\documentclass[a4paper,11pt]{article}
\usepackage{pos}

\title{Quantum Decoherence at ESSnuSB Experiment}
\manuallySeparateAuthors 
\author*[a]{Monojit Ghosh}
\author{ for the ESSnuSB Collaboration}

\affiliation[a]{Center of Excellence for Advanced Materials and Sensing Devices, Ru{\dj}er Bošković Institute, 10000 Zagreb, Croatia}

\emailAdd{mghosh@irb.hr}

\abstract{
In this proceedings we study the sensitivity of the ESSnuSB experiment to probe quantum decoherence. ESSnuSB is a future long-baseline neutrino oscillation experiment which aims to measure $\delta_{\rm CP}$ by probing the second oscillation maximum. Using the open quantum system formalism for decoherence, we have shown that the sensitivity of ESSnuSB to constrain the decoherence parameters is better than MINOS but comparable to DUNE. We have also shown that the CP measurement capability of ESSnuSB is robust in presence of decoherence. }

\FullConference{12th Neutrino Oscillation Workshop (NOW2024)\\
 2-8, September 2024\\
Otranto, Lecce, Italy\\}

\begin{document}
\maketitle

\section{Introduction}

The future long-baseline neutrino oscillation experiment ESSnuSB \cite{Alekou:2022emd} provides a great opportunity to explore several physics scenarios beyond the standard three flavour framework. One of such scenario is quantum decoherence. Neutrino oscillation is a quantum mechanical interference phenomenon which occurs due to the coherent superposition of different neutrino mass eigenstates. When the coherence in the interference pattern is lost, the neutrino mass states becomes decoherent. There are two kinds of decoherence which which are generally studied in the neutrino system: kinematic decoherence and dynamic decoherence. Kinematic decoherence arises in wave packet formalism of the neutrino oscillations which is described by the separation of the wave packets due to different velocities in the mass states. Dynamic Decoherence follows the  open quantum system formalism which is basically environment induced decoherence. In this proceedings, we adopt the open quantum system formalism. In this case, neutrino as a subsystem interacting with the environment give rise to decoherence. The evolution equation of the neutrinos is given by the Lindblad Master equation \cite{Lindblad:1975ef}  

\begin{equation}
	\frac{\partial\rho(t)}{\partial t}=-i[H,\rho(t)] + \mathcal{D}[\rho(t)]\,,
    \label{eq:lindblad}
\end{equation}
where $\rho(t)$ is the density matrix corresponding to the neutrino states and $H$ is the neutrino (subsystem) Hamiltonian in the presence of ambient matter. The effect of decoherence is given by the dissipator matrix  $\mathcal{D}$. This matrix, in the three neutrino case can be expanded as $\mathcal{D}=\mathcal{D}_{jk}\rho_k \lambda_j$, where $\lambda_j$ are the Gell-Mann matrices and $\rho_k$ are the elements of the neutrino density matrix. Finally, the probability of neutrino oscillation in presence of this kind of decoherence takes the following form:
\begin{equation}
	P_{\alpha \beta} = {\rm Tr}\left[\rho_{\alpha} (0) \rho_\beta (t)\right]
    \label{eq:prb}
\end{equation}
Therefore, in order to compute the neutrino oscillation probabilities in presence of decoherence, one needs to consider a particular form of $\mathcal{D}_{jk}$, solve Eq.~\ref{eq:lindblad} for $\rho$ and calculate Eq.~\ref{eq:prb}. 

Regarding the form of $\mathcal{D}_{jk}$, we consider the one that is studied in Ref.~\cite{BalieiroGomes:2018gtd} i.e., 
\begin{equation}
 \label{eq:Dmatrixdiag}
	D_{jk}= -\mathrm{diag}(\Gamma_{21},\Gamma_{21},0,\Gamma_{31}, \Gamma_{31},\Gamma_{32},\Gamma_{32},0)\,.
\end{equation}
with 
\begin{align}
    \Gamma_{31} = \Gamma_{21} + \Gamma_{32} - 2\sqrt{\Gamma_{21}\Gamma_{32}}\,.
    \label{eq:gamma_relation}
\end{align}
This gives rise to the following expression of the neutrino oscillation probabilities:
\begin{align}
    P(\nu_\alpha\to\nu_\beta)=\delta_{\alpha\beta} &- 2\sum_{i>j} {\rm Re}\left[ \tilde{U}^*_{\alpha i}\tilde{U}_{\beta i}\tilde{U}_{\beta j}\tilde{U}^*_{\beta j} \right] ~\left[1-\cos \left(2\tilde{\Delta}_{ij} \right)~e^{-\Gamma_{ij}L}\right]\nonumber
    \\
    &+2\sum_{i>j}{\rm Im}\left[ \tilde{U}^*_{\alpha k}\tilde{U}_{\beta k}\tilde{U}_{\beta j}\tilde{U}^*_{\beta j} \right] ~\sin \left(2\tilde{\Delta}_{ij}\right)~e^{-\Gamma_{ij}L}\,,
    \label{eq:finalprob}
\end{align}
where $\tilde{U}$ is the modified PMNS matrix in matter and $\tilde{\Delta}_{ij} = \dfrac{\Delta\tilde{m}^2_{ij}L}{4E}$, with $\Delta\tilde{m}^2_{ij}$ being the mass squared differences in the presence of matter. Here it is important to note that this formula is only applicable if matter effect is small. $\mathcal{D}$ is conventionally defined in vacuum and therefore inclusion of matter effect requires rotation of $\mathcal{D}$ in matter basis. In this case $\mathcal{D}$ becomes off-diagonal and the probability formula gets altered. This affects neutrinos at higher energies and for this reason the above mentioned formalism remains valid in the context of ESSnuSB where the matter effect is small.

In this proceedings, we will discuss the capability of the ESSnuSB experiment to constrain the two independent parameters of decoherence i.e., $\Gamma_{21}$ and $\Gamma_{32}$ and show their effect in the measurement of $\delta_{\rm CP}$.

\section{The ESSnuSB Experiment}

The current ESSnuSBplus consortium includes 13 countries and 23 institutions. The ESSnuSB experiment will consist of a water Cherenkov far detector of fiducial volume 538 kt,  located at a distance of 360 km at Zinkgruvan mine from the neutrino source in Lund. A powerful linear accelerator (linac) will be used to produce $2.7\times 10^{23}$ protons on target per year with a beam power of 5 MW and proton kinetic energy equal to 2.5 GeV.  In our analysis we have considered a 10 year runtime divided into 5 years in neutrino mode and 5 years in antineutrino mode. Our calculations are done using the software  GLoBES~\cite{Huber:2004ka}.

\section{Results}

\begin{figure}[h]
%\hspace*{-1.75cm}
\includegraphics[width=7cm,height=6.5cm]{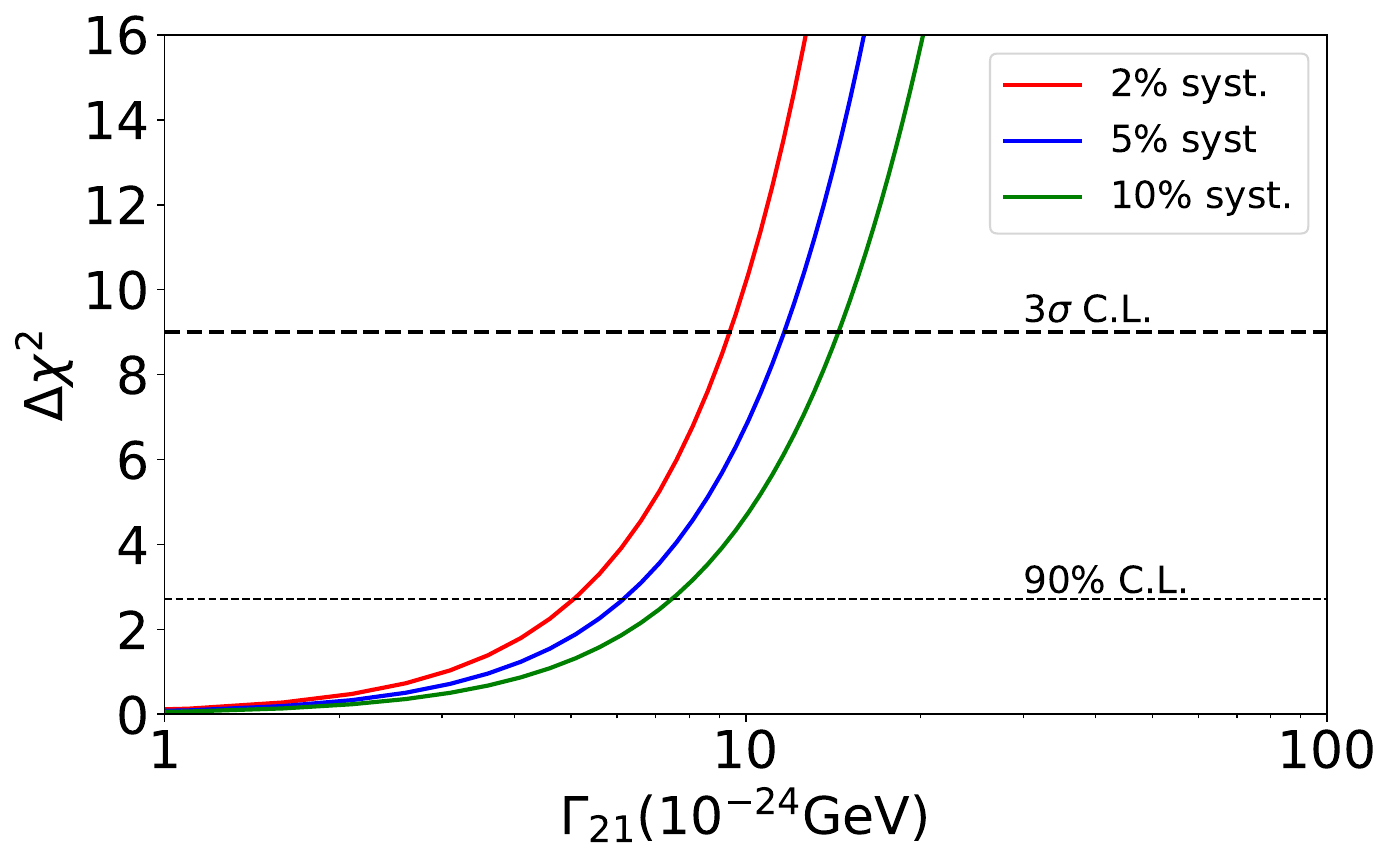}
%\hspace*{0.5cm}
\includegraphics[width=7cm,height=6.5cm]{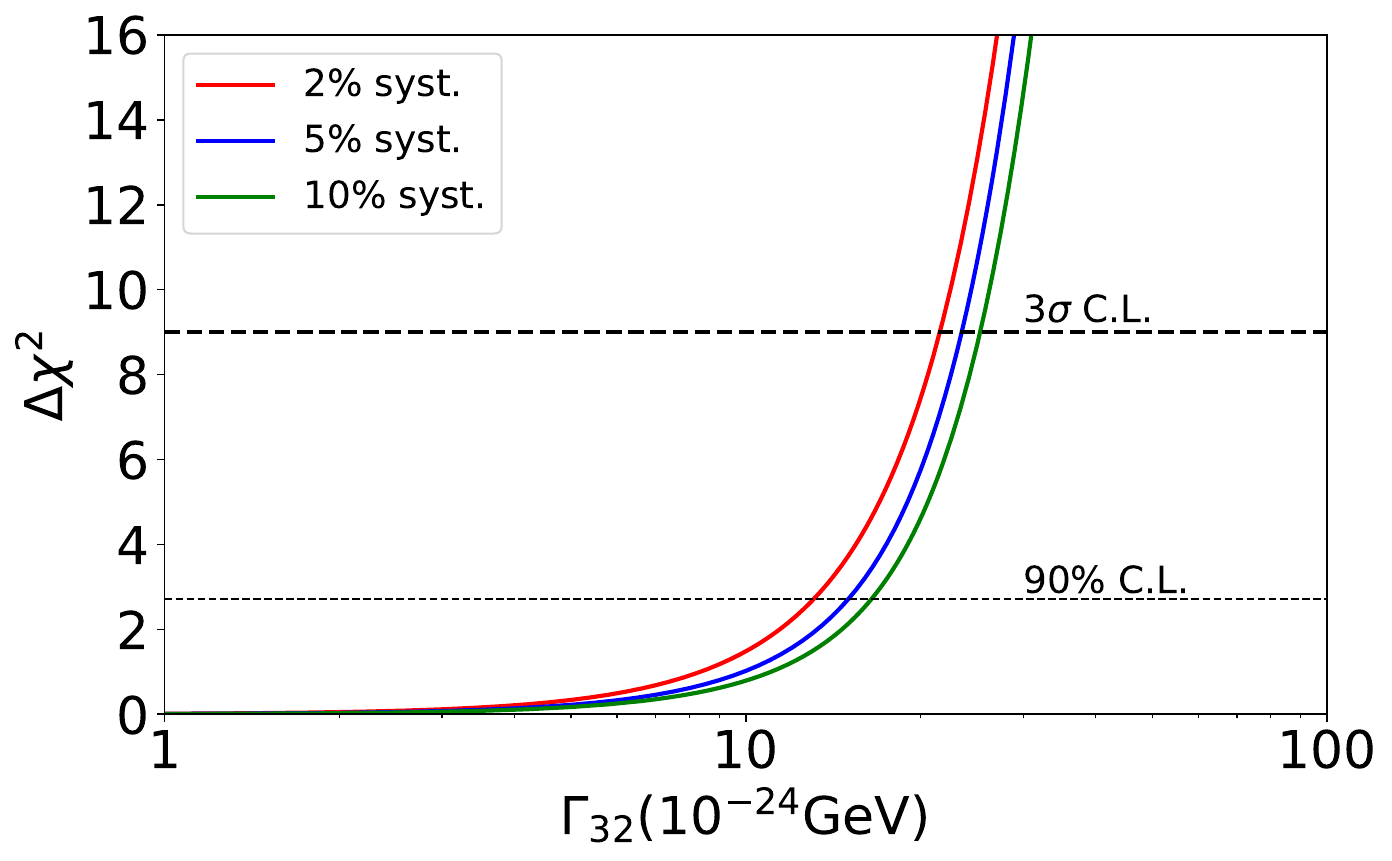}
\caption{Constraints on the decoherence parameters from the ESSnuSB experiment.}
\label{fig:bnd}
\end{figure}
\begin{figure}[h]
\includegraphics[width=5cm,height=4.5cm]{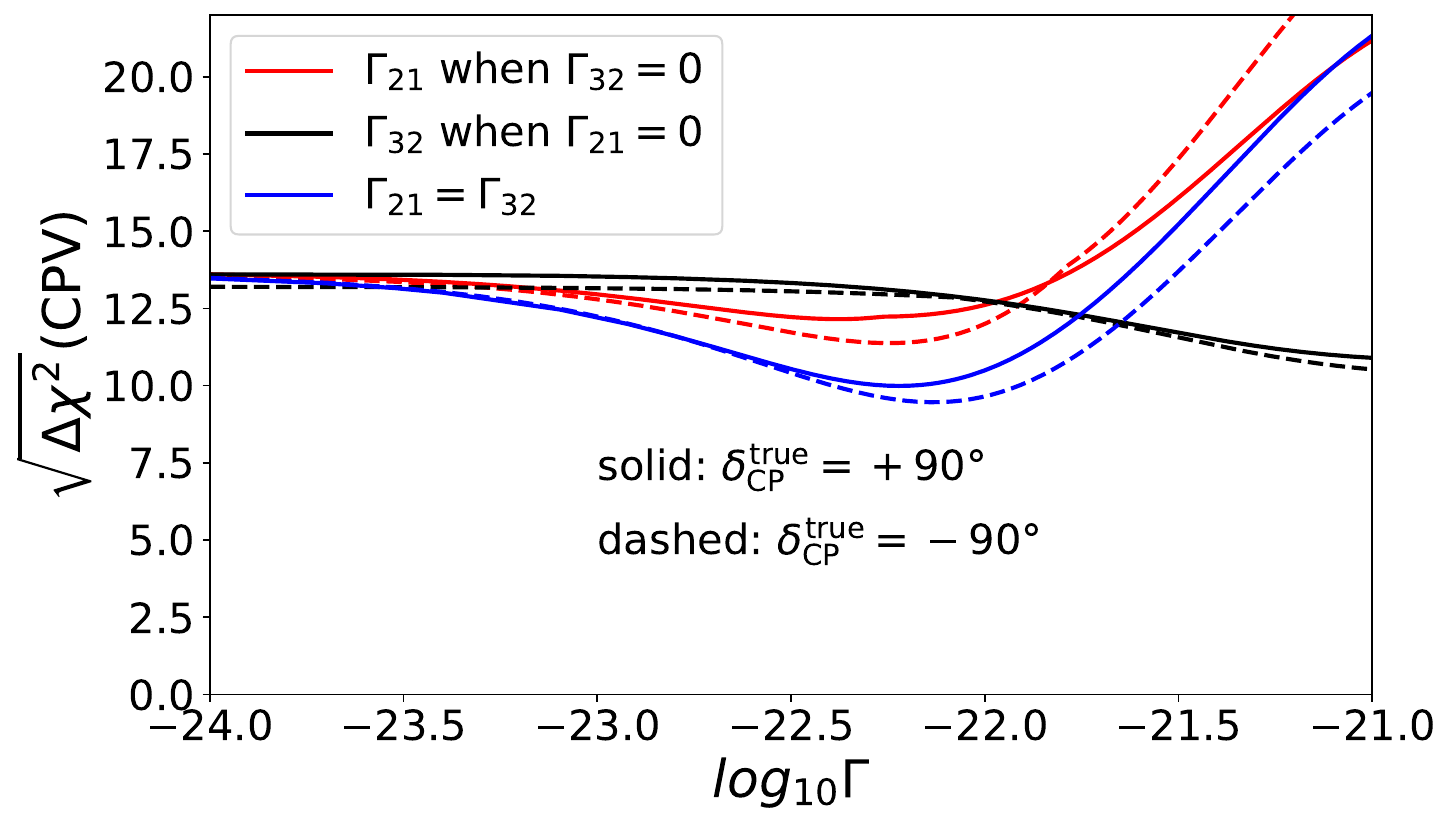}
\hspace*{-0.1 cm}
\includegraphics[width=5cm,height=4.5cm]{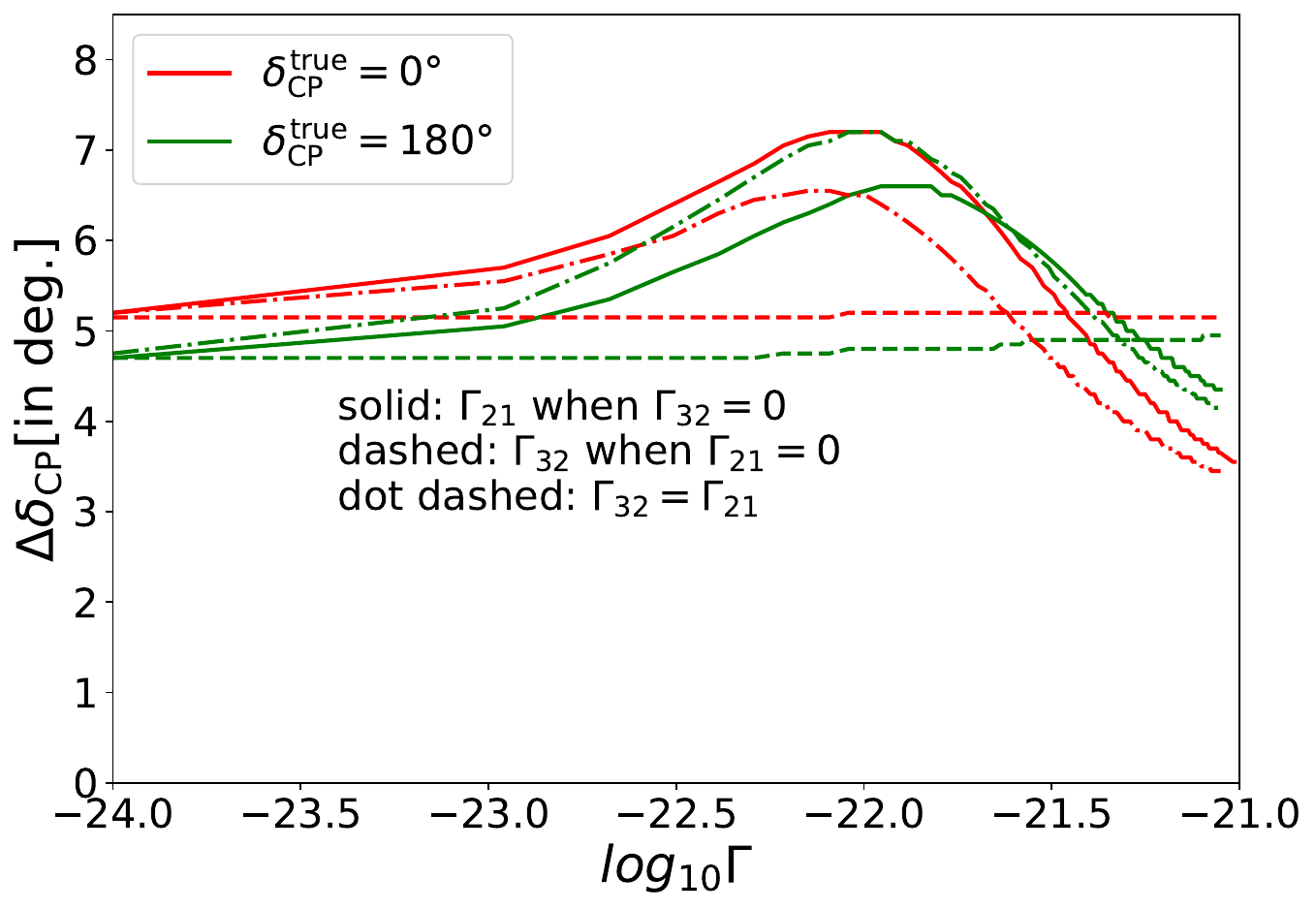}
\hspace*{-0.1 cm}
\includegraphics[width=5cm,height=4.5cm]{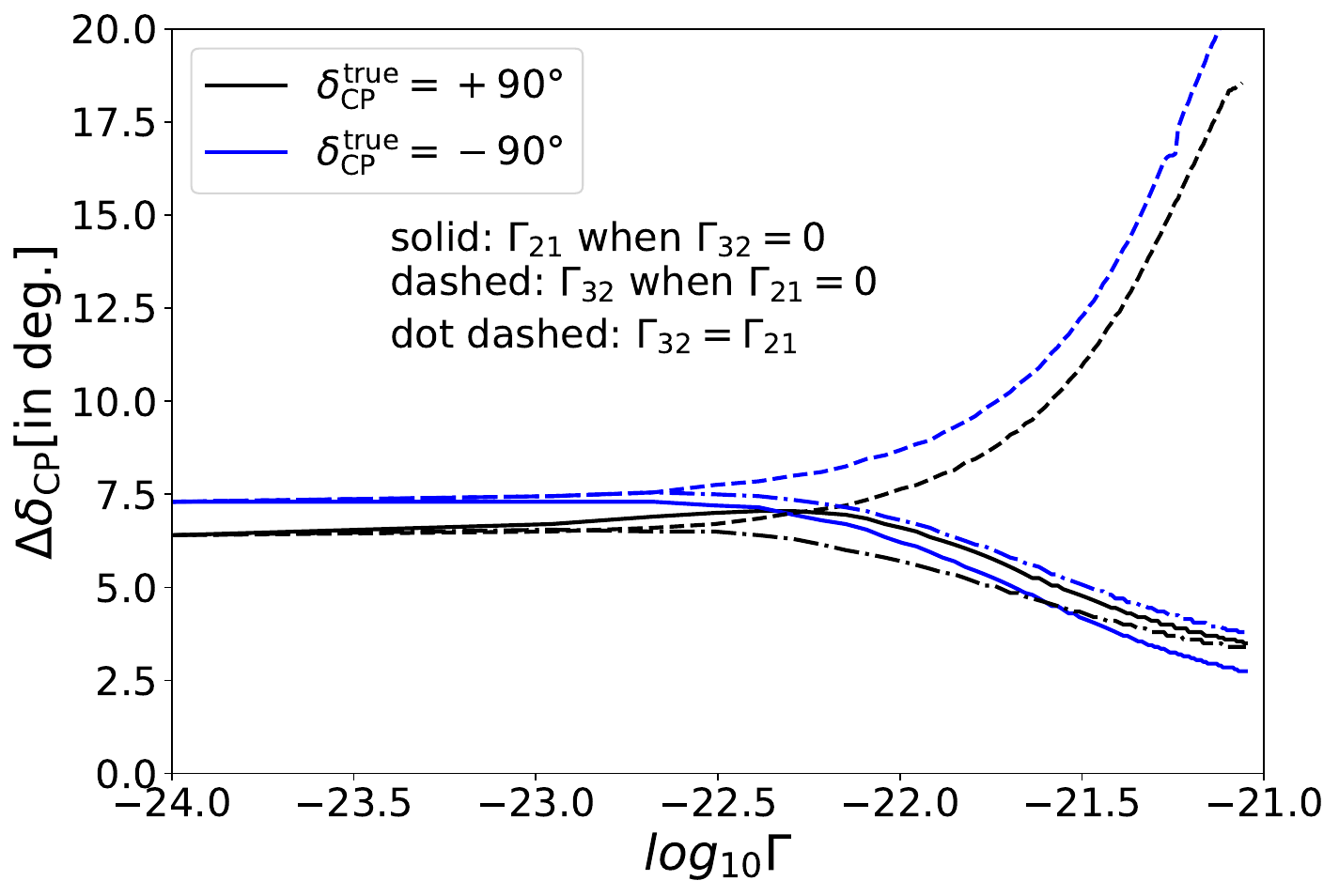}
\caption{Effect of the decoherence parameters in the $\delta_{CP}$ sensitivity.}
 %\caption{}
 \label{fig:cp}
\end{figure}

In Fig.~\ref{fig:bnd}, we have shown the capability of ESSnuSB to put bound on the decoherence parameter. The left panel is for $\Gamma_{21}$ and the right panel is for $\Gamma_{32}$. In each panel we have shown sensitivity for three values of systematic errors. In Fig.~\ref{fig:cp}, we have shown how the CP sensitivity of ESSnuSB is affected in presence of decoherence. The left panel shows the CP violation sensitivity, the middle (right) panel shows the CP precision sensitivity for $\delta_{\rm CP} = 0^\circ$ and $180^\circ$ ($\pm 90^\circ$).

\section{Conclusion}

In this proceedings we have studied quantum decoherence in the context of ESSnuSB experiment. Our results show that the sensitivity is better than MINOS \cite{DeRomeri:2023dht} and comparable to DUNE \cite{BalieiroGomes:2018gtd}.  Further, CP sensitivity is not affected much if the decoherence parameters are not very large. For more details see Ref.~\cite{ESSnuSB:2024yji} on which this proceedings is based upon.

\section*{Acknowledgements}

Funded by the European Union. Views and opinions expressed are however those of the author(s) only and do not necessarily reflect those of the European Union. Neither the European Union nor the granting authority can be held responsible for them.

\end{document}